\documentclass{article}
\usepackage{amssymb}
\usepackage{amsmath,bm,shuffle}
\usepackage{array}
\usepackage{caption,graphicx}
\usepackage[all]{xy}
\newcolumntype{C}[1]{>{\centering\arraybackslash }b{#1}}

\newcommand{\be}{\begin{equation}}
\newcommand{\ee}{\end{equation}} 

\def\un{{\rm 1\mkern-4mu I}}

\title{Deducing the symmetry of the standard model from the automorphism and structure groups of the exceptional Jordan algebra
\footnote{Updated version of IHES/P/17/03 with minor corrections.}}
\author{Ivan Todorov$^1$, Michel Dubois-Violette$^2$,}
\date{
\small 
$^1$Institut des Hautes \'Etudes Scientifiques, 35 route de Chartres, 
\\ F-91440 Bures-sur-Yvette -- France
\\ Institute for Nuclear Research and Nuclear Energy, Bulgarian Academy of Sciences, 
\\ Tsarigradsko Chaussee 72, BG-1784 Sofia -- Bulgaria
\\ (permanent address)\\
$^2$ Laboratoire de Physique Th\'eorique, CNRS, Universit\'e Paris-Sud, 
\\ Universit\'e Paris-Saclay, B\^at. 210, F-91405 Orsay -- France}

\begin{document}

\maketitle

\vglue 2cm

\begin{abstract}
We continue the study undertaken in \cite{DV} of the exceptional Jordan algebra $J = J_3^8$ as (part of) the finite-dimensional quantum algebra in  an almost classical space-time approach to particle physics. Along with reviewing known properties of $J$ and of the associated exceptional Lie groups we argue that the symmetry of the model can be deduced from the Borel-de Siebenthal theory of maximal connected subgroups of simple compact Lie groups.
\end{abstract}

\newpage

%\tableofcontents

\section{Introduction}\label{sec1}

The exceptional Jordan algebra $J = J_3^8 = H_3 ({\mathbb O})$ -- the algebra of $3 \times 3$ hermitian matrices with octonionic entries (reviewed in \cite{McC,J68,B,Be,BS,G,Y}) -- appears to be tailor made for the description of three families of quarks and leptons (like $\begin{pmatrix} u &c &t \\ \nu_e &\nu_{\mu} &\nu_{\tau} \end{pmatrix}$ or $\begin{pmatrix} d &s &b \\ e &\mu &\tau \end{pmatrix}$ of a fixed chirality) -- see \cite{DV}, briefly outlined in Sect.~2. There are three exceptional Lie algebras associated with $J$:
\begin{enumerate}
\item[(a)] the automorphism or {\it derivation algebra}
\be
\label{eq11}
{\rm Der} \, (J) = {\mathfrak f}_4 \, (= {\rm Lie} \, F_4) = so (9) \, \dot + \, S_9 = so (8) \, \dot + \, S_8^+ \, \dot + \, S_8^- \, \dot + V_8,
\ee
(we use, following \cite{BS}, the sign $\, \dot + \, $ for the direct sum of vector spaces, to be distinguished from the direct sum $\oplus$ of (mutually commuting) algebras);
\item[(b)] the (reduced) {\it structure algebra}
\be
\label{eq12}
{\rm str} \, (J) = {\mathfrak e}_6 \, (= {\rm Lie} \, E_6) = {\mathfrak f}_4 \, \dot + \,  J_0 \, ;
\ee
\item[(c)] the {\it conformal algebra}
\be
\label{eq13}
{\rm co} \, (J) = {\mathfrak e}_7 \, (= {\rm Lie} \, E_7) = {\mathfrak e}_6 \, \dot + \, 2J \, \dot + \, {\mathbb C} \, .
\ee
\end{enumerate}

\noindent Here Lie $G$ stands for the Lie algebra of the Lie group $G$; $S_9$ is the 16-dimensional spinor representation of the rotation Lie algebra $so (9)$; it can be viewed as the direct sum of the two inequivalent 8-dimensional spinor representations $S_8^{\pm}$ of $so (8)$; $V_8$ is the (8-dimensional) vector representation of $so (8)$; $J_0$ is the traceless part of $J$ (the 26 dimensional real vector space of $3 \times 3$ hermitian traceless octonionic matrices, also denoted as $sH_3 ({\mathbb O})$). The construction of the above exceptional Lie algebras involves the {\it magic square} of Freudenthal and Tits. It is explained in \cite{BS,B} -- see the summary in Sect.~3.

\smallskip

The Borel-de~Siebenthal theory (see \cite{BdS, K}) describes the maximal closed connected subgroups of a compact Lie group that have maximal rank. Our main observation (Sect.~4) is that the intersection of the maximal subgroups ${\rm Spin} \, (9)$, and 
$\frac{SU (3) \times SU (3)}{{\mathbb Z}_3}$ of the (compact) automorphism group $F_4$ of $J$ is the gauge group of the standard model of particle physics
\be
\label{eq14}
G_{F_4} = G_{\rm ST} = S(U(2) \times U(3)) = \frac{SU (2) \times SU (3) \times U(1)}{{\mathbb Z}_6}.
\ee
This result makes it natural to consider as possible extensions of $G_{\rm ST}$ the intersections of appropriate maximal rank subgroups of ${\rm Str} \, (J) = E_6$,
\be
\label{eq15}
G_{E_6} = S(U(2) \times U(2) \times U(3)) 
\ee
and of ${\rm Co} (J) = E_7$;
\be
\label{eq16}
G_{E_7} = S(U(2) \times U(3) \times U(3)) \, .
\ee
Note that all three groups, $G_{F_4}$, $G_{E_6}$, $G_{E_7}$ are non-semisimple (i.e. they include $U(1)$ factors) compact subgroups of $F_4 , E_6 , E_7$, of maximal rank ($4,6,7$, respectively).

\smallskip

It would be useful to consider $J$ as a member of the family $H_n ({\mathbb K})$ where ${\mathbb K}$ is an {\it alternative composition algebra}. We recall that an algebra ${\mathcal A}$ is said to be a {\it division algebra} if $ab=0$ for $a,b \in {\mathcal A}$ implies that either $a=0$ or $b=0$. It is called an {\it alternative algebra} if any two elements of ${\mathcal A}$ generate an associative subalgebra. Zorn has proven (in 1933) that there are just four  alternative division algebras: the real and the complex numbers, ${\mathbb R}$ and ${\mathbb C}$, the quaternions, ${\mathbb H}$, and the octonions, ${\mathbb O}$. All four admit a multiplicative norm $x \to \vert x \vert \in {\mathbb R}_+$ such that
\be
\label{eq17}
\vert xy \vert = \vert x \vert \, \vert y \vert \, , \quad \vert x \vert^2 = x \, \bar x = \bar x \, x \quad (>0 \ {\rm for} \ x \ne 0)
\ee
where $x \to \bar x$ is the (involutive) conjugation in ${\mathbb K}$. Hurwitz has proven back in 1898 that the only normed division algebras are ${\mathbb R}$, ${\mathbb C}$, ${\mathbb H}$ and ${\mathbb O}$. $H_n ({\mathbb K})$ is the algebra of $n \times n$ hermitian matrices (with entries in ${\mathbb K}$) closed under the Jordan multiplication
\be
\label{eq18}
X \circ Y = \frac12 (XY + YX) \, (= Y \circ X) \, .
\ee
For ${\mathbb K} = {\mathbb O}$ the resulting algebra only satisfies the {\it Jordan condition}
\be
\label{eq19}
(X^2 \, Y) X = X^2 (YX)
\ee
for $n =1,2,3$. The condition (\ref{eq19}), on the other hand, characterizes an abstract {\it Jordan algebra} for which the endomorphism $Z \to X(YZ) - Y(XZ)$ is an (inner) derivation.

\smallskip

Division algebras can be also characterized by the existence of a non-degene\-rate real trilinear form $t : {\mathbb K}^{\times 3} \to {\mathbb R}$, the {\it triality form} -- see Sect.~2.4 of \cite{B}. (We note that it looks nontrivial even for ${\mathbb K} = {\mathbb C}$. In this case, for $z_j = x_j + i \, y_j$, $j=1,2,3$, the form is a multiple of $t(z_1 , z_2 , z_3) = x_1 \, x_2 \, x_3 - x_1 \, y_2 \, y_3 - x_2 \, y_1 \, y_3 - x_3 \, y_1 \, y_2$.) In general, the presence of a hermitian inner product (cf. (\ref{eq17})) makes the existence of $t$ equivalent to the existence of a ${\mathbb K}$-valued cross product in ${\mathbb K}$.

\smallskip

We shall consider (in Sect.~2 and Sect.~3.3) more general {\it alternative composition algebras} which have a (not necessarily positive definite) non-degenerate sesquilinear form $\langle x,y \rangle$ satisfying the (square of the) factorization property (\ref{eq17}):
\be
\label{eq110}
\langle xy , xy \rangle = \langle x,x \rangle \, \langle y,y \rangle \, ,
\ee
and have an alternating associator (see Eq. (\ref{eq23}) below).

\section{Exceptional finite quantum geometry}\label{sec2}
\setcounter{equation}{0}

In the approach of almost commutative geometry [DKM, D,  CL, C, CC, CCS, BF] to the standard model,  space-time is viewed as the tensor product of a standard (commutative) 4-dimensional spin manifold with a finite noncommutative space. In the almost classical quantum geometry approach one is led to identify the finite quantum space with the exceptional Jordan algebra.\\

\smallskip

To begin with, it was argued in \cite{DV} that the decomposition of the (8-dimensional, real) vector space ${\mathbb O}$ of octonions\footnote{Octonios have been first applied to the standard model by Feza G\"ursey and his students \cite{G}. His work triggered an imaginative development by G. Dixon \cite{Di} followed by C. Furey \cite{Fu} among others. A distinguished feature of our approach, started in \cite{DV}, is the fact that we are dealing with an euclidean Jordan algebra suited for a (finite) \textit{observable} algebra.} into a direct sum of complex vector spaces,
\be
\label{eq21}
{\mathbb O} = {\mathbb C} \oplus {\mathbb C}^3
\ee
naturally corresponds to the splitting of the basic fermions (in one generation) of the standard model into quarks and leptons. Moreover, the color group $SU(3)$ leaves invariant a (complex) volume form on ${\mathbb C}^3$ which is dual with respect to the hermitian scalar product $\langle \ , \ \rangle$ in ${\mathbb C}^3$ to a skew symmetric antilinear cross product ${\bf z} \times {\bf w} = - \, {\bf w} \times {\bf z}$ ($\times : {\mathbb C}^3 \otimes {\mathbb C}^3 \to {\mathbb C}^3$): $Vol({\bf z_1}, {\bf z_2}, {\bf z_3})=<{\bf z_1}\times{\bf z_2}, {\bf z_3}>$. This cross product and the (hermitian) inner product $\langle {\bf z,w} \rangle$ ($\langle \ , \ \rangle : {\mathbb C}^3 \otimes {\mathbb C}^3 \to {\mathbb C}$) can be combined with the usual multiplication of complex numbers and extended to a unique real bilinear multiplication $xy$ in ${\mathbb O}$ that is norm preserving: 
$$
\langle xy , xy \rangle = \langle x,x \rangle \, \langle y,y \rangle \, ,
$$
\be
\label{eq22}
x = z_0 \oplus {\bm z} \Rightarrow \langle x,x \rangle = \bar z_0 \, z_0 + \sum_{i=1}^3 \bar z_i \, z_i \ (\equiv \vert x \vert^2)
\ee
($\bar z$ standing for the complex conjugate of $z \in {\mathbb C}$). The resulting product is non associative but {\it alternative}: the {\it associator}
\be
\label{eq23}
[x,y,z] = (xy)z - x(yz)
\ee
is an alternating function of $x,y,z$; in particular, it vanishes if any two of the arguments $x,y,z$ coincide. More generally, this is true for any {\it composition algebra} -- i.e. an algebra with a non-degenerate (but not necessarily positive definite -- thus including the {\it split octonions}) inner product $\langle \ , \ \rangle$ satisfying the first equation (\ref{eq22}) (cf. \cite {BS}).

\smallskip

The significance of the notion of an alternative algebra is illustrated by the following remark. The commutator
$$
ad_a \, x = [a,x]
$$
defines a {\it derivation} in an associative algebra:
$$
ad_a \, (xy) = (ad_a \, x) \, y + x \, ad_a \, y \, .
$$ 
This property fails, in general, for a non-associative algebra. If however the algebra ${\mathcal A}$ is {\it alternative}, every pair of elements $x,y \in {\mathcal A}$ defines a derivation $D_{x,y}$ in ${\mathcal A}$ in terms of the double commutator and the associator:
\be
\label{eq24}
D_{x,y} (z) = \left[[x,y] , z \right] - 3 \, [x,y,z]
\ee
(see Eq.~(14) of \cite{B}).

\smallskip

The construction of the octonionic scalar product satisfying (\ref{eq22}) in terms of the cross product and the inner product in ${\mathbb C}^3$, indicated above (and worked out in \cite{DV, TD}) yields the standard multiplication in ${\mathbb O}$ which is conveniently expressed in terms of the Fano plane of imaginary octonionic units recalled in Appendix~A. Choosing, say, $e_7$ as the ``$i$'' in ${\mathbb C}$ we can write the decomposition (\ref{eq21}) explicitly in the basis $\{e_0 = 1 , e_j , j = 1,\ldots , 7 \}$ as:
$$
x = x^0 + x^7 \, e_7 + (x^1 + x^3 \, e_7) \, e_1 + (x^2 + x^6 \, e_7) \, e_2 \, +
$$
\be
\label{eq25}
+ \, (x^4 + x^5 \, e_7) \, e_4 \left( = \sum_{\alpha = 0}^7 x^{\alpha} \, e_{\alpha} \right) .
\ee

The presence of three generations of quark-lepton doublets $\begin{pmatrix} u &\nu \\ d &e \end{pmatrix}$ (with $u,d$ -- 3-vectors in the color space) suggests combining the octonions into a $3 \times 3$ hermitian matrix:
\be
\label{eq26}
X = \begin{pmatrix}
\xi_1 &x_3 &\bar x_2 \\
\bar x_3 &\xi_2 &x_1 \\
x_2 &\bar x_1 &\xi_3 \end{pmatrix} , \quad x_i \in {\mathbb O} \, , \quad \xi_i = \xi_i \in {\mathbb R}
\ee
where the bar over an octonion $x$ stands for octonionic conjugation (changing simultaneously the sign of all imaginary units $e_j$, $j=1,\ldots ,7$).

\smallskip

The matrices (\ref{eq26}) span a 27-dimensional real vector space which can be given the structure of the exceptional Jordan algebra $J = J_3^8 \, (= H_3 ({\mathbb O}))$ with multiplication defined as the symmetrized matrix multiplication (\ref{eq18}):
\be
\label{eq27}
X \circ Y = \frac12 \, (XY + YX) \, .
\ee
As emphasized in \cite{DV} the (axiomatic) properties of the (commutative) Jordan product $\circ$ are dictated by the requirement to have a spectral decomposition for (hermitian) observables. In fact, the requirement of {\it formal reality} ($\underset{i}{\sum} \, x_i^2 = 0 \Rightarrow x_i = 0$ for all $i$) implies that the {\it Jordan condition} (\ref{eq19})
\be
\label{eq28}
(x^2 \circ y) \circ x = x^2 \circ (y \circ x)
\ee
is equivalent to power associativity 
\be
\label{eq29}
x^r x^s = x^{r+s} \qquad (x \in J \, , \ r,s \in {\mathbb N})
\ee
as proven in \cite{JvNW} (see Theorem 1 of \cite{DV}). It is clearly necessary for the standard theory of spectral decomposition.

\smallskip

We recall the remark after Eq.~(\ref{eq19}) according to which the Jordan condition (\ref{eq28}) ensures that the commutator of two left multiplications is a derivation:
\be
\label{eq210}
[L_x , L_y] \in {\rm Der} \, J \quad \mbox{for} \quad L_x (y) := x \circ y \, , \ x,y \in J
\ee
(i.e. $[L_x , L_y] \, (z \circ w) = [L_x , L_y] \, (z) \circ w + z \circ [L_x , L_y] \, (w)$).

\smallskip

One also defines a real linear function ${\rm tr} \, X$, a bilinear inner product $\langle X,Y \rangle$ and a symmetric trilinear form ${\rm tr} \, (X,Y,Z)$ on $J$ setting
$$
{\rm tr} \, X = \xi_1 + \xi_2 + \xi_3 \, (= \langle X , \un \rangle) \, ,
$$
$$
\langle X,Y \rangle = {\rm tr} \, (X \circ Y) \, ,
$$
\be
\label{eq211}
{\rm tr} \, (X,Y,Z) = \langle X , Y \circ Z \rangle = \langle X \circ Y , Z \rangle \, .
\ee

\section{The Lie algebra of derivations of $J$ and its extensions: the structure and the conformal algebras}\label{sec3}
\setcounter{equation}{0}

There are excellent detailed expositions of the material of this section. We share the opinion of John Baez \cite {B} that to survey the early developments of this subject ``one still cannot do better than to read Freudenthal's classic 1964 paper \cite{F} on Lie groups and foundations of geometry''. Later work including the 1966 Vinberg's and the 1976 Ramond's (triality) constructions is given a self contained treatment in \cite{BS} (appearing about the same time as Baez's temperamental survey). The less emotional 2009 review by Yokota \cite{Y} provides a systematic treatment of exceptional Lie groups (with all formulas needed to follow the details). The present short survey aims to fix our notation and to formulate the results that will be used in Sect.~4.

\subsection{The automorphism group $F_4$ of the exceptional Jordan algebra and its Lie algebra ${\mathfrak f}_4$}

About the same time Pascual Jordan introduced his algebras Ruth Moufang studied her non-Desarguian (octonionic) projective plane $P{\mathbb O}_2$. Sixteen years later, in 1949, Jordan noticed that the points of $P{\mathbb O}_2$ are given by the one-dimensional (trace-one) idempotents of the exceptional Jordan algebra $J$ which are also the pure states of $J$.\\
A glimpse on the automorphism group  $F_4 = \{ g : J \to J ; g(X \circ Y) = g \, X \circ g \, Y \}$ is provided by displaying the stability subgroup of one such idempotent
\be
\label{eq31}
E_1 = \begin{pmatrix}
1 &0 &0 \\
0 &0 &0 \\
0 &0 &0 \end{pmatrix} \left( {\rm or} \ E_2 = \begin{pmatrix}
0 &0 &0 \\
0 &1 &0 \\
0 &0 &0 \end{pmatrix} , \ {\rm or} \ E_3 = \begin{pmatrix}
0 &0 &0 \\
0 &0 &0 \\ \, 
0 &0 &1 \end{pmatrix} \right).
\ee
Noting that $F_4$ should preserve the unit element $\un = E_1 + E_2 + E_3$ of $J$ we deduce that this stability subgroup should also preserve $E_2 + E_3$ and hence the square of any traceless element of $H_2 ({\mathbb O})$,
\be
\label{eq32}
X^2 = (\xi^2 + \vert x \vert^2) \begin{pmatrix}
0 &0 &0 \\
0 &1 &0 \\
0 &0 &1 \end{pmatrix} \ \mbox{for} \quad X = \begin{pmatrix}
0 &0 &0 \\
0 &\xi &x \\
0 &\bar x &-\xi \end{pmatrix}
\ee
and coincides with ${\rm Spin} \, (9)$ (the simply connected double covering of the orthogonal group $SO (9)$ in nine dimensions). It follows that the octonionic projective (\textit{Moufang}) plane\footnote{The octonionic quantum mechanics in the Moufang plane is studied in \cite{GPR}.} coincides with the homogeneous space
\be
\label{eq33}
{\mathbb P} {\mathbb O}_2 = F_4 / {\rm Spin} (9) \, .
\ee
This allows to find, in  particular, the dimension of $F_4$ (over the reals):
\be
\label{eq34}
\dim \, (F_4) = \dim \, ({\rm Spin} (9)) + \dim \, ({\mathbb P} {\mathbb O}_2) = 36 + 16 = 52 \, .
\ee
With a little more work one recovers the Lie algebra ${\mathfrak f}_4$ as a direct sum of the Lie algebra $so (9)$ and its 16-dimensional spinor representation $S_9$:
\be
\label{eq35}
{\mathfrak f}_4 \simeq so (9) \, \dot + \, S_9 \qquad (S_9 = S_8^+ \, \dot + \, S_8^-)
\ee
which yields (\ref{eq11}) and can be interpreted in ``purely octonionic'' terms:
\be
\label{eq36}
{\mathfrak f}_4 \cong so \, ({\mathbb O} \oplus {\mathbb R}) \, \dot + \, {\mathbb O}^2 = so \, ({\mathbb O}) \, \dot + \, {\mathbb O}^3 \, .
\ee

Finally, we turn to a description that will also apply to higher rank exceptional Lie algebras. According to \cite{BS} the Lie algebra of derivations on $H_3 ({\mathbb K})$ -- the set of hermitian $3 \times 3$ matrices over any alternative composition algebra ${\mathbb K}$ -- can be presented as a sum
\be
\label{eq37}
{\rm Der} \, (H_3 ({\mathbb K})) \cong {\rm Der} \, ({\mathbb K}) \, \dot + \, sa_3 ({\mathbb K})
\ee
where $sa_3 ({\mathbb K})$ is the set of antihermitean traceless $3 \times 3$ matrices with entries in ${\mathbb K}$:
\be
\label{eq38}
sa_3 ({\mathbb K}) = \{ X \in {\mathbb K} \, [3] ; \ X^* = -X , \ {\rm tr} \, X = 0 \} \, .
\ee
Given an element $X \in sa_3 ({\mathbb K})$ there is a derivation $ad_X$ of $H_3 ({\mathbb K})$ given by
\be
\label{eq39}
ad_X (Y) = [X,Y] \quad \mbox{for} \quad \forall \, Y \in H_3 ({\mathbb K}) \, .
\ee
The subspace ${\rm Der} \, ({\mathbb K})$ in the right-hand side of (\ref{eq37}) is always a Lie algebra, but $sa_3 ({\mathbb K})$ is not unless ${\mathbb K}$ is commutative and associative (in which case ${\rm Der} \, ({\mathbb K})$ vanishes). Nevertheless, there is a formula for the bracket in ${\rm Der} \, [H_3({\mathbb K})]$ which applies in every case. Given $D,D' \in {\rm Der} \, [H_3({\mathbb K})]$ and $X,Y \in sa_3 ({\mathbb K})$ it reads:
$$
[D,D'] = DD' - D'D \, , \quad [D,ad_X] = ad_{DX} \, ,
$$
\be
\label{eq310}
[ad_X , ad_Y] = ad_{[X,Y]_0} + \frac13 \sum_{i,j=1}^3 D_{x_{ij} , y_{ij}}
\ee
where $D$ acts on
$$
X = \begin{pmatrix}
x_{11} &x_{12} &x_{13} \\
x_{21} &x_{22} &x_{23} \\
x_{31} &x_{32} &x_{33} 
\end{pmatrix}
$$
componentwise, $[X,Y]_0$ is the trace-free part of the commutator $[X,Y]$, and $D_{x,y}$ is the derivation defined by (\ref{eq24}).

\smallskip

Summarizing, we have the following expressions for the compact form of ${\mathfrak f}_4$ (which also appears as the isometry algebra of the Riemannian manifold ${\mathbb P} {\mathbb O}^2$):
\be
\label{eq311}
{\mathfrak f}_4 \cong {\rm Der} \, H_3 ({\mathbb O}) \cong {\rm Der} \, ({\mathbb O}) \, \dot + \, sa_3 ({\mathbb O}) 
\ee
(a special case of (\ref{eq37})). Here ${\rm Der}\, ({\mathbb O})$ is the 14-dimensional exceptional Lie algebra $\mathfrak g_2$.

\subsection{The magic square}

Equivalent constructions of the Lie algebras ${\mathfrak e}_n$ with $n  = 6,7,8$ have been proposed by Freudenthal and Tits around 1958, with improved formulations published later. In the summary below we follow \cite{BS}, as well as \cite{B} where more references to the early work can be found.

\smallskip

Let ${\mathbb K}$ be a real composition algebra and $J$ a real Jordan algebra with unit $\un$ and with an inner product satisfying $\langle X , Y \circ Z \rangle = \langle X \circ Y , Z \rangle$ (\ref{eq210}). Let further ${\mathbb K}_0$ and $J_0$ be the subspaces of ${\mathbb K}$ and $J$ orthogonal to the unit element. Denote by $*$ the product in $J_0$ obtained from the Jordan product projected back to $J_0$:
\be
\label{eq312}
X * Y = X \circ Y - \frac1n \, \langle X,Y \rangle \, \un \quad \mbox{where} \quad n = \langle \un , \un \rangle = {\rm tr} \, \un
\ee
(the notation being chosen to fit the case $J = H_n ({\mathbb K}')$). Tits defines (in 1966) a Lie algebra structure on the vector space
\be
\label{eq313}
T ({\mathbb K} , J) = {\rm Der} \, ({\mathbb K}) \, \dot + \, {\rm Der} \, (J) \, \dot + \, {\mathbb K}_0 \otimes J_0
\ee
by setting
\be
\label{eq314}
[x \otimes X , y \otimes Y] = \frac1n \, \langle X,Y \rangle \, D_{x,y} - \langle x,y \rangle [L_X , L_Y] + [x,y] \otimes X * Y
\ee
where $x,y \in {\mathbb K}_0$, $X,Y \in J_0$ and the square brackets in the right hand side denote commutators in ${\mathbb K}_0$ and ${\rm End} \, (J)$; $D_{x,y}$ is the derivation in ${\mathbb K}_0$ defined by (\ref{eq24}). Tits proves that the brackets (\ref{eq314}) define a Lie algebra structure using the identity
\be
\label{eq315}
n \, X^3 - ({\rm tr} \, X^3) \, \un = ({\rm tr} \, X^2) \, X \quad \mbox{for} \quad X \in J_0 \, , \ J = H_n ({\mathbb K}')
\ee
(for ${\mathbb K}' = {\mathbb O}$, $n \leq 3$). Tits obtains the magic square of Lie algebras by viewing $T({\mathbb K} , J)$ for $J = H_3 ({\mathbb K}')$ as a Lie algebra $L({\mathbb K} , {\mathbb K}')$ depending on two composition algebras ${\mathbb K}$ and ${\mathbb K}'$:
\be
\label{eq316}
L({\mathbb K} , {\mathbb K}') = T ({\mathbb K} , H^3 ({\mathbb K}')) \, .
\ee
For the Lie algebras of compact real forms one thus obtains the following symmetric table:

\bigskip

\begin{tabular}{|C{1cm}|C{2cm}|C{2cm}|C{2cm}|C{2cm}|}
\hline ${\mathbb K} \ \backslash \ {\mathbb K}'$ &${\mathbb R}$ &${\mathbb C}$ &${\mathbb H}$ &${\mathbb O}$  \\
\hline ${\mathbb R}$ &$so \, (3)$ &$su \, (3)$ &$sp \, (6)$ &${\mathfrak f}_4$  \\
\hline ${\mathbb C}$ &$su \, (3)$ &$su \, (3) \oplus u(3)$ &$su \, (6)$ &${\mathfrak e}_6$ \\
\hline ${\mathbb H}$ &$sp \, (6)$ &$su \, (6)$ &$so \, (12)$ &${\mathfrak e}_7$ \\
\hline ${\mathbb O}$ &${\mathfrak f}_4$ &${\mathfrak e}_6$ &${\mathfrak e}_7$ &${\mathfrak e}_8$ \\
\hline
\end{tabular}

\smallskip

\begin{quotation}
\noindent {\bf Table 1.} Magic square of Lie algebras $L({\mathbb K} , {\mathbb K}')$ ($sp \, (6)$ being the rank 3 unitary symplectic Lie algebra).
\end{quotation}

\smallskip

Following Tits construction the symmetry of the square comes as a surprise. In fact, it has been predicted in a non-rigorous visionary 1956 paper of the Russian mathematician and historian of science Boris Rosenfeld who proposed to view $E_6 , E_7 , E_8$ as isometry groups of projective planes over the algebras ${\mathbb K} \, \otimes \, {\mathbb O}$ for ${\mathbb K} = {\mathbb C} , {\mathbb H}, {\mathbb O}$, respectively, just as $F_4$ is the isometry group of ${\mathbb P}^2 ({\mathbb O})$ $=$ ${\mathbb P}^2 ({\mathbb R} \, \otimes \, {\mathbb O})$ (see \cite{B} for references and for a more detailed 
discussion; Rosenfeld provides a later expositon of his views in Chapter VII of \cite{R}). The realization of this idea has problems since ${\mathbb K} \otimes {\mathbb O}$ is not a division algebra except for ${\mathbb K} = {\mathbb R}$. A construction of the exceptional Lie algebras generalizing (\ref{eq311}), however, does exist with ${\rm Der} \, ({\mathbb K}) \oplus {\rm Der} \, ({\mathbb O})$ instead of ${\rm Der} \, ({\mathbb O})$ and $sa_3 ({\mathbb O})$ substituted by $sa_3 ({\mathbb K} \otimes {\mathbb O})$. This is Vinberg's (1966) approach to constructing $L({\mathbb K} , {\mathbb K}')$, that is manifestly symmetric with respect to the two algebras ${\mathbb K}$ and ${\mathbb K}'$. The Lie brackets in $L({\mathbb K} , {\mathbb K}')$ are given as follows.

\begin{enumerate}
\item[(i)] ${\rm Der} \, ({\mathbb K})$ and ${\rm Der} \, ({\mathbb K}')$ are commuting Lie subalgebras of $L ({\mathbb K} , {\mathbb K}')$.
\item[(ii)] The bracket of $D \in {\rm Der} \, ({\mathbb K}) \oplus {\rm Der} \, ({\mathbb K}')$ with $X \in sa_3 ({\mathbb K} \otimes {\mathbb K}')$ is given by applying $D$ to every entry of the matrix $X$ using the natural action of ${\rm Der} \, ({\mathbb K})$ and ${\rm Der} \, ({\mathbb K}')$ as derivations on ${\mathbb K} \otimes {\mathbb K}'$.
\item[(iii)] Given $X,Y \in sa_3 ({\mathbb K} \otimes {\mathbb K}')$, we set
\be
\label{eq317}
[X,Y] = [X,Y]_0 + \frac13 \sum_{i,j=1}^3 D_{x_{ij} , y_{ij}} \, .
\ee
Here $[X,Y]_0$ is the traceless part of the $3 \times 3$ matrix $[X,Y]$, and given $x,y \in {\mathbb K} \otimes {\mathbb K}'$, we define $D_{x,y} \in {\rm Der} \, ({\mathbb K}) \oplus {\rm Der} \, ({\mathbb K}')$ as real bilinear in $a,b \in {\mathbb K}$, $a',b' \in {\mathbb K}'$ such that
\be
\label{eq318}
D_{a \otimes a' \, b \otimes b'} = \langle a',b' \rangle \, D_{a,b} + \langle a,b \rangle \, D_{a',b'}
\ee
where $a,b \in {\mathbb K}$, $a',b' \in {\mathbb K}'$, and $D_{a,b}$, $D_{a',b'}$ are defined as in Eq.~(\ref{eq24}).
\end{enumerate}

\smallskip

For the equivalence of Tits' and Vinberg's constructions of the magic square and for its triality construction we refer to \cite{B,BS}.

\subsection{The exceptional Lie groups $E_6$ and $E_7$}

A non-compact real form of the simply connected Lie group $E_6$, the (reduced) {\it structure group} of the exceptional Jordan algebra $J=H_3 ({\mathbb O})$, can be defined as the group of determinant preserving linear transformations of $J$ where, for $X$ given by (\ref{eq26}),
\begin{eqnarray}
\det X &= &\frac13 \, {\rm tr} (X^3) - \frac12 \, {\rm tr} X^2 \, {\rm tr} X + \frac16 \, ({\rm tr} X)^3 \\
&= &\xi_1 \, \xi_2 \, \xi_3 - \xi_1 \vert x_1 \vert^2 - \xi_2 \vert x_2 \vert^2 - \xi_3 \vert x_3 \vert^2 + 2 {\rm Re} \, x_1 \, x_2 \, x_3 \, . \nonumber
\end{eqnarray}
Noting that the Lie algebra of this non-compact group has the form (\ref{eq12}) ${\rm str} \, (J) = {\mathfrak f}_4 \, \dot + \, J_0$ where ${\mathfrak f}_4$ is, in fact, its maximal compact Lie subalgebra, one finds that the signature of the Killing form of ${\rm str} \, (J)$ is\footnote{We are using the common notation (cf. \cite{B}); \cite{BS} write instead ${\mathfrak e}_6(26)$.} 
\be
\label{eq320}
\mbox{signature} \ [{\rm str} \, (J)] = \dim \, (J_0) - \dim \, {\mathfrak f}_4 = 26-52 = -26 \, .
\ee
It is demonstrated in \cite{BS} that this non-compact form of ${\mathfrak e}_6$ is obtained if we replace the complex numbers ${\mathbb C}$ in $L({\mathbb C} , {\mathbb O})$ by the split alternative algebra $\widetilde{\mathbb C}$:
\be
\label{eq321}
{\rm Str} \, J = E_{6 (-26)} \, , \quad {\mathfrak e}_{6 (-26)} = L (\widetilde{\mathbb C} , {\mathbb O})
\ee
where the {\it split form} $\widetilde{\mathbb C}$ of ${\mathbb C}$ is obtained by replacing the {\it imaginary unit} $i$ by $e$ such that $e^2-1 = (e+1)(e-1)=0$ ($\widetilde{\mathbb K}$ is {\it split} if at least one of the ``imaginary units'' has square $1$). (Actually, ${\mathfrak e}_{6(-26)}$ is identified with ${\rm str}' (J)$ in \cite{BS}, the prime indicating factorization with respect to the multiples of the central operator $L_{\un}$ (of left multiplication by the unit element in ${\mathbb K}$).)

\smallskip

The asymmetrical Tits construction of the above Lie algebra gives
\be
\label{eq322}
{\mathfrak e}_6 \cong {\rm Der} \, (H_3 ({\mathbb O})) + sH_3 ({\mathbb O}) = {\mathfrak f}_4 + J_0
\ee
where $sH_3 ({\mathbb K})$ stands for the traceless hermitian $3 \times 3$ matrices with entries in ${\mathbb K}$. Eq.~(\ref{eq322}) allows to easily calculate the dimension of ${\mathfrak e}_6$ :
\be
\label{eq323}
\dim {\mathfrak e}_6 = \dim \, ({\mathfrak f}_4) + \dim \, (J_0) = 52+26 = 78 \, .
\ee
As it is demonstrated in \cite{A} ${\mathfrak e}_6$ can be decomposed as a vector space into the maximal rank Lie subalgebra $so (10) \oplus u(1)$ and the 32 dimensional space $S_{10}$ of $so (10)$ spinors:
\be
\label{eq324}
{\mathfrak e}_6 = so (10) \oplus u(1) \, \dot + \, S_{10} \, ;
\ee
moreover the natural mapping $S_{10} \times S_{10} \to so (10)$ allows to reconstruct the Lie bracket in the compact form $e_6$.

\smallskip

In 1954 Freudenthal described a non-compact form $E_{7 (-25)}$ of $E_7$ as a group of linear transformations of the 56-dimensional space ${\mathcal P}$ of block matrices
\be
\label{eq325}
P = \begin{pmatrix}
\alpha &X \\
Y &\beta \end{pmatrix} \quad \mbox{where} \quad \alpha , \beta \in {\mathbb R} \, , \ X,Y \in J \quad (\alpha \equiv \alpha \, \un_3 \, , \ \beta \equiv \beta \, \un_3)
\ee
that preserve the symplectic form
\be
\label{eq325bis}
\omega \, (P_1 , P_2) = \alpha_1 \, \beta_2 - \alpha_2 \, \beta_1 + \langle X_1 , Y_2 \rangle - \langle Y_1 , X_2 \rangle
\ee
and a triple product ${\mathcal P}^{\times 3} \to {\mathcal P}$. The maximal compact subgroup of $E_{7(-25)}$ is $E_6 \times U(1) / {\mathbb Z}_3$, the Lie algebra $e_{7(-25)}$ having a vector space decomposition
\be
\label{eq325ter}
{\rm co} \, (J) = e_{7(-25)} = e_6 \oplus u(1) \, \dot + \, J \, \dot + \, J
\ee
of signature
\be
\label{eq326}
\mbox{signature} \ [{\rm co} \, (J)] = 2 \dim \, (J) - \dim \, [e_6 \oplus u(1)] = 54-79 = -25 
\ee
justifying the above notation.

\smallskip

A review of the Kantor-Koecher-Tits construction of ${\rm co} (J)$ that explores the correspondence between a {\it Jordan triple system} and 3-graded Lie algebras is contained in \cite{P}.

\smallskip

The Tits construction of the compact form of ${\mathfrak e}_7$, on the other hand, yields the elegant relation
\begin{eqnarray}
\label{eq327}
{\mathfrak e}_7 = {\rm Der} \, (H_3 ({\mathbb O})) \, \dot + \, [H_3 ({\mathbb O})]^3 \Rightarrow \dim \, ({\mathfrak e}_7) &= &\dim \, ({\mathfrak f}_4) + 3 \, \dim \, (J) \nonumber \\
&= &52+81=133 \, .
\end{eqnarray}

\newpage

\section{Borel-de Siebenthal theory and intersections of maximal subgroups of compact exceptional Lie groups}\label{sec4}
\setcounter{equation}{0}

Borel and de Siebenthal \cite{BdS} described the maximal maximal-rank subgroups of simple compact connected Lie groups noticing that each such subgroup appears as the identity component of the centralizer of its center. This yields the following explicit classification of the maximal subalgebras of the simple compact Lie algebras:
\begin{eqnarray}
\label{eq41}
su (n+1) &: &su(p+1) \times su(n-p) \times u(1) \quad (p=1 , \ldots , \left[ \frac n2 \right]); \nonumber \\
so (2n+1) &: &so(2n) \, , \ so (2p+1) \times so (2n-2p) \, , \ so (2n-1) \times u(1) \, , \nonumber \\
&&(p=1,\ldots , n-2); \nonumber \\
sp (2n) &: &sp(2p) \times sp (2n-2p) \, , \ su (n) \times u(1) \quad (p=1, \ldots , \left[ \frac n2 \right]); \nonumber \\
so (2n) &: &so (2p) \times so (2n-2p) \, , \ so (2n-2) \times u(1) \, , \ su (n) \times u(1); \nonumber \\
e_6 &: &su (2) \times su (6) \, , \ su (3) \times su (3) \times su (3) \, , \ so (10) \times u(1); \nonumber \\
e_7 &: &su (2) \times so (12) \, , \ su (3) \times su (6) \, , \ su(8) \, , \ e_6 \times u(1); \nonumber \\
e_8 &: &so(16) \, , \ su (9) \, , \ su (5) \times su (5) \, , \ e_6 \times su (3) \, , \ e_7 \times su (2); \nonumber \\
{\mathfrak f}_4 &: &so(9) \, , \ su (3) \times su (3) \, , \ su (2) \times sp (6) ; \nonumber \\
g_2 &: &su (3) \, , \ su (2) \times su (2) \, .
\end{eqnarray}

\noindent Baez and Huerta \cite{BH} have observed that the intersection of the grand unified theory (GUT) symmetry groups $SU (5)$ (of Georgi-Glashow) and $\frac{{\rm Spin} \, (4) \times {\rm Spin} \, (6)}{{\mathbb Z}_2}$ ($= \, \frac{SU(2) \times SU(2) \times SU(4)}{{\mathbb Z}_2}$ of Pati-Salam) within the ${\rm Spin} \, (10)$ grand unification (also introduced by Georgi in 1974) coincides with the gauge group $G_{\rm ST}$ (\ref{eq14}) of the standard model. We shall see that the Borel-de Siebenthal theory provides a purely deductive path to this gauge group.

\smallskip

We first note that the intersection  of the maximal connected subgroups of $G_2$ is
\be
\label{eq42}
G_{G_2} = SU(3) \cap \frac{SU (2) \times SU (2)}{{\mathbb Z}_2} = U(2)
\ee
-- that is, the gauge group for the Weinberg-Salam model. This however destroys our rational for introducing the octonions: the unbroken $SU(3)$ color symmetry. We shall therefore restrict the maximal subgroups of $G_2$ and of the higher rank groups under consideration. A maximal rank subgroup of an exceptional Lie group will be called \textit{admissible} if it contains the color $SU(3)$ subgroup. Thus the only admissible subgroup of $G_2$ is $SU(3)(=SU(3)_c)$ itself.

Assuming that the exceptional Jordan algebra $J$ is a good candidate for the finite geometry underlying the standard model of particle physics, it would be natural to view its automorphism group $F_4$ as (its possible) GUT symmetry. If we then look for the intersection of its maximal admissible subgroups (described by Theorems 2.9.1 and 2.12.2 (but excluding 2.11.2!) of \cite{Y}):
\be
\label{eq43}
{\rm Spin} \, (9) \, , \quad \frac{SU (3) \times SU (3)}{{\mathbb Z}_3},  
\ee
one finds precisely the group $G_{\rm ST} = G_{F_4}$ (\ref{eq14}):
\be
\label{eq44}
G_{F_4} = S[U(2) \times U(3)] = \frac{SU (3) \times SU (2) \times U(1)}{{\mathbb Z}_6} \, . 
\ee
 Similarly, for the compact form $E_6$ of the (reduced) structure group of $J$ we find (according to theorems 3.10.7,  and 3.13.5 of \cite{Y} - the subgroup described by 3.11.4 being inadmissible):
\begin{eqnarray}
\label{eq45}
G_{E_6} &= &\frac{{\rm Spin} \, (10) \times U (1)}{{\mathbb Z}_4} \cap \frac{SU (3) \times SU (3) \times SU(3)}{{\mathbb Z}_3} \nonumber \\
&= &S[U(2) \times U(2) \times U(3)] \, , 
\end{eqnarray}
a group with an extra $U(1)$ factor and a remnant of the Pati-Salam model that is favored in \cite{CCS} (see also \cite{BF}). Finally, for the compact form of $E_7$ we find  (using 4.10.2, 4.11.15 and 4.13.5 and excluding  4.12.5 of \cite{Y}):
\begin{eqnarray}
\label{eq46}
G_{E_7} &= &\frac{E_6 \times U (1)}{{\mathbb Z}_3} \cap \frac{{\rm Spin} \, (12) \times SU (2)}{{\mathbb Z}_2} \cap \frac{SU (6) \times SU (3)}{{\mathbb Z}_3} \nonumber \\
&= &S(U(2) \times U(3) \times U(3)) \, .
\end{eqnarray}

Since the early work of G\"ursey, Ramond and Sikivie \cite{GRS} one uses the 27-dimensional representation of $E_6$ to combine one generation of fermions (the {\bf 16} of ${\rm Spin} \, (10) \subset E_6$) with the bosonic representations $\bm{10} \oplus {\bm 1}$ of ${\rm Spin} \, (10)$. A similar interpretation is given to the basic representation  $\underline{26}$ of $F_4$ in
\cite{TD}. The interpretation of either of the groups (\ref{eq43}) (\ref{eq45}) and (\ref{eq46}) will depend on the choice of representation of the exceptional Jordan algebra. As pointed out in \cite{DV} any finite module over $J$ is isomorphic to $J \otimes E$ for some finite dimensional vector space $E$.  It was argued in (Sect.~4.4 of) \cite{DV}, another attractive candidate for a finite quantum algebra may be
\be
\label{eq47}
J_1^2 \oplus J_2^4 \oplus J_3^8 = {\mathbb R} \oplus H_2 ({\mathbb H}) \oplus H_3 ({\mathbb O}) \, .
\ee
We leave the study of these possibilities and their physical implications to future work.

\smallskip

\noindent {\bf Acknowledgments.}  I.T. thanks for hospitality IHES where the bulk of this work was done and the Theoretical Physics Department of CERN where the paper was completed. The work of I.T. has been supported in part by Project DN 18/1 of the Bulgarian National Science Foundation.

\newpage

\section*{Appendix A. The Fano plane of imaginary octonions (\cite{B})}\label{secAA}

$$
\includegraphics[width=4cm]{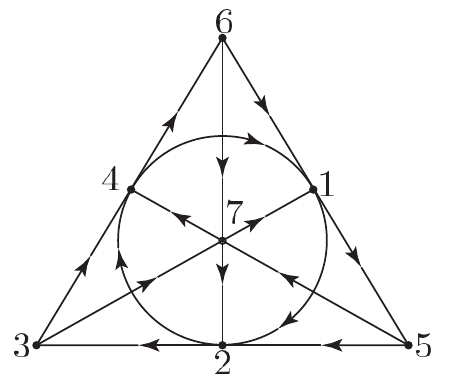}
$$
$$
e_1 = (0,0,1) , e_2 = (0,1,0) \Rightarrow e_1 \, e_2 = e_4 = (0,1,1)
$$
$$
e_3 = (1,0,0) \Rightarrow e_2 \, e_3 = e_5 = (1,1,0)
$$
$$
e_1 \, e_5 = e_6 = (1,1,1)
$$
$$
e_4 \, e_5 = e_7 = (1,0,1) \, .
$$

\centerline{\bf Figure 1.}

\smallskip

\centerline{
\noindent Projective plane in ${\mathbb Z}_2^3$ with seven points and seven  lines.
}

\bigskip

The multiplication table for the seven octonionic imaginary units can be recovered from the following properties:
$$
e_i^2 = -1 \, , \quad i=1,\ldots , 7 \, ; \quad e_i \, e_j = -e_j \, e_i \, ; \eqno ({\rm A}.1)
$$
$$
e_i \, e_j = e_k \Rightarrow e_{i+1} \, e_{j+1} = e_{k+1} \, , \quad e_{2i} \, e_{2j} = e_{2k} \eqno ({\rm A}.2)
$$
where indices are counted modulo seven; and a single relation of the type
$$
e_1 \, e_2 = e_4 \eqno ({\rm A}.3)
$$
producing a quaternionic line. We have displayed on Fig.~1 the points $e_i$ as non-zero triples of homogeneous coordinates taking values $0$ and $1$ such that the product $e_i \, e_j$ (in clockwise order) is obtained by adding the coordinates $(a,b,c)$, $a,b,c \in \{0,1\}$, modulo two.

\end{document}